\documentstyle[twocolumn,seceq]{jpsj}
\input{epsf.tex}

\newcommand \im{{\rm i}}
\newcommand \D{\,{\rm d}}

\newcommand \sech{{\rm sech}}

\newcommand \dn{{\rm dn}}
\newcommand \rhs{r.\,h.\,s.}
\newcommand \lhs{l.\,h.\,s.}

\def\runtitle { Integrable Magnetic Model of Two Chains Coupled by
  Four-Body Interactions }

\def\runauthor { Norihiro {\sc Muramoto} and Minoru {\sc Takahashi} }

\title { Integrable Magnetic Model of Two Chains \\ Coupled by
  Four-Body Interactions }

\author { Norihiro {\sc Muramoto} and Minoru {\sc Takahashi} }

\inst { Institute for Solid State Physics, University of Tokyo,
  Roppongi, Minato-ku, Tokyo 106 }

\recdate { \today }

\abst { An exact solution for an {\it XXZ} chain with four-body
  interactions is obtained and its phase diagram is determined.  The
  model can be reduced to two chains coupled by four-body
  interactions, and it is shown that the ground state of the two-chain 
  model is magnetized in part.  Furthermore, a twisted four-body
  correlation function of the anti-ferromagnetic Heisenberg chain is
  obtained. }

\kword { quantum inverse scattering method, algebraic Bethe ansatz,
  thermodynamic Bethe ansatz }

\begin{document}
\sloppy \maketitle

\section{Introduction}
Quantum spin systems solved exactly so far are essentially one
dimensional.  It is a challenging problem to find a two, or higher,
dimensional exactly solvable quantum spin model.  As a first step we
study exactly solvable models for two, or more, chains combined each
other by some interactions.  A well-known example is the
Majumdar-Ghosh model~\cite{MG}, which can be interpreted as a two-leg
zigzag ladder;
\begin{eqnarray}
  \label{H:MGmodel}
  H_{\rm MG}
  &=& \frac{1}{2}\sum_{j=1}^{N} \mib{S}_{j}\cdot\mib{S}_{j+1}
    + \frac{1}{2}\sum_{j=1}^{N} \mib{T}_{j}\cdot\mib{T}_{j+1}
  \nonumber\\
  && \mbox{}
    + \sum_{j=1}^{N} \mib{S}_{j}\cdot\mib{T}_{j}
    + \sum_{j=1}^{N} \mib{T}_{j}\cdot\mib{S}_{j+1},
\end{eqnarray}
where $\mib{S}_{j}$ and $\mib{T}_{j}$ are operators of
spin-$\frac{1}{2}$ at site $j$ on each chain.  The ground states are
given by doubly degenerate dimer states with no magnetization.

We introduce here a novel two-chain model, which is exactly solvable
by the Bethe ansatz method.  The model consists of two Heisenberg
chains coupled by four-body interactions.  It can be expressed as
twisted two chains;
\begin{eqnarray}
  \label{H:2chain}
  H_{\rm 2chain}
  &=& \frac{1}{4}\sum_{j=1}^{N} \mib{S}_{j}\cdot\mib{S}_{j+1}
    + \frac{1}{4}\sum_{j=1}^{N} \mib{T}_{j}\cdot\mib{T}_{j+1}
  \nonumber\\
  && \mbox{}
  + \sum_{j=1}^{N}
  \left( \mib{S}_{j  }\!\times\!\mib{T}_{j  } \right)\cdot
  \left( \mib{S}_{j+1}\!\times\!\mib{T}_{j+1} \right)\nonumber\\
  && \mbox{}
  + \sum_{j=1}^{N}
  \left( \mib{T}_{j  }\!\times\!\mib{S}_{j+1} \right)\cdot
  \left( \mib{T}_{j+1}\!\times\!\mib{S}_{j+2} \right).
\end{eqnarray}
This Hamiltonian is obtained as a special case of the following
single-chain model;
\begin{eqnarray}
  \label{H:single}
  H_{\Delta,\alpha}
  &=&
  J_{1} \sum_{j=1}^{N}
  [\mib{S}_{j}\cdot\mib{S}_{j+1}]_{\Delta_{1}}
  +
  J_{2} \sum_{j=1}^{N}
  \mib{S}_{j}\cdot\mib{S}_{j+2}
  \nonumber\\
  &+&
  J_{3} \sum_{j=1}^{N}
  \Bigl(
    (\mib{S}_{j  }\cdot\mib{S}_{j+2})
    (\mib{S}_{j+1}\cdot\mib{S}_{j+3})
  \Bigr.
  \nonumber\\
  && \mbox{}
  -\Bigl.
     [\mib{S}_{j  }\cdot\mib{S}_{j+3}]_{\Delta_{3}}
     [\mib{S}_{j+1}\cdot\mib{S}_{j+2}]_{\Delta_{3}'}
  \Bigr).
  \nonumber\\
  &+&
  {\rm const.},
\end{eqnarray}
where an abbreviation of anisotropic scalar product is used,
\begin{equation}
  \label{abbre}
  [\mib{a}\cdot\mib{b}]_{\Delta}
  \equiv
  a^{x}b^{x} + a^{y}b^{y} + \Delta a^{z}b^{z}.
\end{equation}
This model obtains integrability when the anisotropies, $\Delta_{1}$,
$\Delta_{3}$, and $\Delta'_{3}$, and the strengths of interactions,
$J_{k}$, satisfy certain conditions with two parameters, $\Delta$ and
$\alpha$.

In the next section the integrability condition of the single-chain
model~(\ref{H:single}) is presented.  The model is integrable if its
Hamiltonian is decomposed to a sum of mutually commuting operators
including the {\it XXZ} model Hamiltonian.  The eigenvalues of the
commuting operators are evaluated in \S3.  The existence of phase
transition is discussed in \S4, where the phase boundary is determined 
analytically.  The magnetic property of the two-chain model is
mentioned briefly in \S5.  In \S6 a correlation function of the
nearest four sites of the Heisenberg chain is explicitly given.  The
last section is devoted to a brief summary.

\section{Integrability}

We consider the system of the following Hamiltonian;
\begin{equation}
  \label{H:def}
  H_{\Delta,\alpha} = H_{\it XXZ} + \alpha H_{\rm int},
\end{equation}
where the first term on the {\rhs} is the Hamiltonian of the usual
anti-ferromagnetic {\it XXZ} model,
\begin{equation}
  \label{H:XXZ}
  H_{\it XXZ}
  =
  \sum_{j=1}^{N} [\mib{S}_{j}\cdot\mib{S}_{j+1}]_{\Delta}
  - \frac{\Delta}{4}N,
\end{equation}
and the interaction term, $H_{\rm int}$, involves up to the third
neighbor interactions;
\begin{eqnarray} 
  \label{H:int}
  \!\!\!\!\!\!\!
  H_{\rm int}
  &=&
  -\frac{1+\Delta^2}{2}\sum_{j=1}^{N}
  [\mib{S}_{j}\cdot\mib{S}_{j+1}]_{\frac{2\Delta}{1+\Delta^2}}
  \nonumber\\
  &+&
  \frac{\Delta}{2}\sum_{j=1}^{N}
  \mib{S}_{j}\cdot\mib{S}_{j+2}
  \nonumber\\
  &+&
  2\Delta \sum_{j=1}^{N}
  \Bigl(
    (\mib{S}_{j  }\cdot\mib{S}_{j+2})
    (\mib{S}_{j+1}\cdot\mib{S}_{j+3})
  \Bigr.
  \nonumber\\
  && \mbox{}
  -
  \Bigl.
  [\mib{S}_{j  }\cdot\mib{S}_{j+3}]_{\Delta}
  [\mib{S}_{j+1}\cdot\mib{S}_{j+2}]_{\Delta^{-1}}
  \Bigr)
  +\frac{\Delta}{8}N.
\end{eqnarray}
The coupling constant, $\alpha$, takes a real value, the anisotropy
parameter is restricted in the range $\Delta>-1$, and the periodic
boundary condition, $S_{N+1}=S_1$, is imposed.  It is easy to see that 
the Hamiltonian~(\ref{H:def}) has a similar form to that of the
single-chain model~(\ref{H:single}).  The former can be considered to
be a special case of the latter.

Note that the parameter $\alpha$ does not denote the strength ratio of 
the second-neighbor, or four-body, interaction to the nearest neighbor
one, but expresses only the ratio between $H_{\rm int}$ and $H_{\it
  XXZ}$.  The strengths, $J_{k}$, and the anisotropies, $\Delta_{k}$,
of the interactions in the Hamiltonian (\ref{H:single}) should be
related with the two parameters, $\alpha$ and $\Delta$, as follow;
\begin{eqnarray}
  \label{eqn:ratio1}
  J_{1} &=& 1 - \frac{1+\Delta^2}{2}\alpha,\;\;\;
  J_{2}  =  \frac{\Delta\alpha}{2},\;\;\;
  J_{3}  =  2\Delta\alpha,\;\;\;
  \\
  \label{eqn:ratio2}
  \Delta_{1} &=& \frac{2\Delta(1-\alpha)}{2-(1+\Delta^2)\alpha},\;\;\; 
  \Delta_{3}  = (\Delta_{3}')^{-1} = \Delta.
\end{eqnarray}
The strength ratio of the second-neighbor interaction to the four-body 
one, $J_{2}/J_{3}$, is fixed at 1/4, though $J_{2}/J_{1}$ has freedom.

Conversely, the Hamiltonian (\ref{H:single}) is integrable under
these conditions, (\ref{eqn:ratio1}) and (\ref{eqn:ratio2}), since it
is decomposed to a sum of two commuting Hamiltonians, (\ref{H:XXZ})
and (\ref{H:int}).  The reason of the commutativity is that the
complicated interaction operator $H_{\rm int}$ is introduced by the
help of the quantum inverse scattering method~\cite{Korepin,Baxter} as
follows.

The Hamiltonian of the {\it XXZ} model is related with the transfer
matrix, $\tau(\lambda)$, of the six-vertex model.  The transfer matrix 
is given as the trace of the monodromy matrix;
\begin{equation}
  \label{def:TM}
  \tau(\lambda) = {\rm tr}\,T(\lambda).
\end{equation}
The monodromy matrix is constructed of $L$-matrices;
\begin{equation}
  \label{def:MM}
  T(\lambda)
  = L_{N}(\lambda)L_{N-1}(\lambda)\cdots L_{1}(\lambda).
\end{equation}
The $L$-matrices are defined as follow;
\begin{eqnarray}
  \label{def:LM}
  \!\!\!\!\!
  L_{j}(\lambda)
  &=& a(\lambda) \frac{1+\sigma_{\rm aux}^{z}\!\otimes\!\sigma_{j}^{z}}{2}
  +   d(\lambda) \frac{1-\sigma_{\rm aux}^{z}\!\otimes\!\sigma_{j}^{z}}{2} 
  \nonumber\\
  &+&
  \frac{1}{\sinh{\frac{\gamma}{2}(\lambda+\im)}}
  \bigl(
    \,\sigma_{\rm aux}^{+}\!\otimes\!\sigma_{j}^{-}
    + \sigma_{\rm aux}^{-}\!\otimes\!\sigma_{j}^{+}
  \bigr),
\end{eqnarray}
with
\begin{eqnarray}
  \label{func:a}
  a(\lambda) &=& \frac{1}{\im\sin\gamma},
  \\
  \label{func:d}
  d(\lambda)
  &=&
  \frac{1}{\im\sin\gamma}
  \frac{\sinh{\frac{\gamma}{2}(\lambda-\im)}}
       {\sinh{\frac{\gamma}{2}(\lambda+\im)}},
\end{eqnarray}
where the parameter $\gamma$ takes a real or a purely imaginary value, 
and $\sigma^{\pm} = \frac{1}{2}(\sigma^{x}\pm\sigma^{y})$.  The
subscript of the Pauli matrices, ${\mib\sigma} = 
(\sigma^{x},\sigma^{y},\sigma^{z})$, denotes the space upon which they 
act.  The trace and the products of $L$-matrices in the definition of
the transfer matrix should be performed in the auxiliary space of the
operator ${\mib\sigma}_{\rm aux}$.

The $L$-matrices satisfy the Yang-Baxter relation;
\begin{equation}
  \label{eqn:YBE}
  R(\lambda,\mu) \Bigl( L(\lambda)\otimes L(\mu) \Bigr)
  = \Bigl( L(\mu)\otimes L(\lambda) \Bigr) R(\lambda,\mu),
\end{equation}
where the $R$-matrix is given as
\begin{eqnarray}
  \label{def:RM}
  R(\lambda,\mu)
  &=& f(\lambda,\mu) \frac{1+\sigma^{z}\!\otimes\!\sigma^{z}}{2}
    + g(\lambda,\mu) \frac{1-\sigma^{z}\!\otimes\!\sigma^{z}}{2}
  \nonumber\\
  && \mbox{}
    + \sigma^{+}\!\otimes\!\sigma^{-}
    + \sigma^{-}\!\otimes\!\sigma^{+},
\end{eqnarray}
with
\begin{eqnarray}
  \label{func:f}
  f(\lambda,\mu)
  &=&
  \frac{\sinh\frac{\gamma}{2}(\lambda-\mu+2\,\im)}
       {\sinh\frac{\gamma}{2}(\lambda-\mu)},
  \\
  \label{func:g}
  g(\lambda,\mu)
  &=&
  \frac{\im\sin\gamma}{\sinh\frac{\gamma}{2}(\lambda-\mu)}.
\end{eqnarray}
Then the transfer matrix forms a commuting family with the spectral
parameter $\lambda$;
\begin{equation}
  \label{eqn:commTM}
  [\tau(\lambda),\tau(\mu)] = 0.
\end{equation}

The $k$-th order Hamiltonian, $H^{(k)}$, is defined as a logarithmic
derivative of the transfer matrix;
\begin{equation}
  \label{H:Hk}
  H^{(k)}
  =
  \left.
    \im\left( \frac{\sin\gamma}{\gamma}
              \frac{\partial}{\partial\lambda} \right)^{k}
    \ln\tau(\lambda)
  \right|_{\lambda=\im}.
\end{equation}
The {\it XXZ} model Hamiltonian~(\ref{H:XXZ}) is identical to the
Hamiltonian of the first order;
\begin{equation}
  H_{\it XXZ} = H^{(1)},\;\;\;
  \Delta = \cos\gamma.
\end{equation}

Because of the commutativity of the transfer matrix, the Hamiltonians
of arbitrary order also commute with one another;
$[H^{(j)},H^{(k)}]=0$.  Then they have common eigenstates, which are
especially common to those of the {\it XXZ} model.  A linear
combination of the commuting Hamiltonians,
\begin{equation}
  H = \sum_{k=1} \alpha_{k} H^{(k)},
\end{equation}
is, therefore, integrable by the Bethe ansatz method.

Tsvelik~\cite{Tsvelik} and Frahm~\cite{Frahm} investigated a model
constructed of the first and the second order Hamiltonian;
\begin{equation}
  \label{H:TF}
  H = H_{\it XXZ} + \alpha H^{(2)},
\end{equation}
where the explicit form of the interaction operator is given as
\begin{eqnarray}
  \label{H:intTF}
  H^{(2)}
  &=&
  \Delta\sum_{j=1}^{N}
  [\mib{S}_{j}\cdot(\mib{S}_{j-1}\!\times\!\mib{S}_{j+1})]_{\Delta^{-1}}.
  \end{eqnarray}
They showed that this model exhibits a phase transition at some finite
value of the coupling constant.  The critical value, $\alpha_{\rm
  crit}$, for arbitrary anisotropy is analytically obtained by Frahm.
The ground state at $|\alpha| > \alpha_{\rm crit}$ obtains an
incommensurability and it is also magnetized in the Ising-like region;
$\Delta\ge1$.

The model, however, has the strange interaction~(\ref{H:intTF}) which
breaks the parity invariance.  That motivated us to choose the new
interaction~(\ref{H:int}), which is nothing but the third order
Hamiltonian;
\begin{equation}
  H_{\rm int} = H^{(3)}.
\end{equation}
This operator conserves the parity.  Furthermore, it contains
isotropic second-neighbor two-body interactions.

When we take the parameter as $\alpha=1$ and $\Delta=1$ in the
Hamiltonian~(\ref{H:def}), the nearest neighbor interactions
cancel out.  It becomes the following form;
\begin{eqnarray}
  \label{H:11}
  \!\!\!\!\!\!\!
  H_{1,1}
  &=&
  \frac{1}{2}\sum_{j=1}^{N} \mib{S}_{j}\cdot\mib{S}_{j+2}
  \nonumber\\
  &+&
  2\sum_{j=1}^{N}
  (\mib{S}_{j  }\!\times\!\mib{S}_{j+1})\cdot
  (\mib{S}_{j+2}\!\times\!\mib{S}_{j+3})
  - \frac{N}{8}.
\end{eqnarray}
This is equivalent with the Hamiltonian of the two-chain model;
$H_{1,1} = 2H_{\rm 2chains}-N/8$.

\section{Eigenstates and Eigenvalues}

In this section we investigate the system of the
Hamiltonian~(\ref{H:def}) by using the algebraic Bethe ansatz
method~\cite{Korepin} and the thermodynamic Bethe ansatz
method~\cite{TakaSuzu,Taka,Gaudin,Taka97,YY}.

We are interested in the ground state of the system.  Then we consider 
the low temperature limit to observe the ground state.

Since the system has the same eigenstates as those of the {\it XXZ}
model, the states with $S_{\rm total}^{z} = (N-2M)/2$ are determined
by the Bethe ansatz equations of the {\it XXZ} model.  Using the
algebraic Bethe ansatz method, we obtain the following equations;
\begin{equation}
  \label{eqn:BAE}
  \left(
    \frac{a(\lambda_{j})}{d(\lambda_{j})}
  \right)^N
  =
  \prod_{l=1,l\neq j}^{M}
  \frac{f(\lambda_{j},\lambda_{l})}{f(\lambda_{l},\lambda_{j})},
  \;\;\;
  j = 1,2,\dots,M.
\end{equation}
where the functions $a(\lambda)$, $d(\lambda)$, and $f(\lambda,\mu)$
are given in the eqs.~(\ref{func:a}), (\ref{func:d}) and
(\ref{func:f}).  The functions on the {\lhs}, $(a(\lambda))^N$ and
$(d(\lambda))^N$, are the eigenvalues of the diagonal elements of the
monodromy matrix, $[T(\lambda)]_{1,1}$ and $[T(\lambda)]_{2,2}$,
respectively, corresponding to the pseudovacuum state;
$|\,\emptyset\,\rangle = \prod_{j=1}^{N}|\!\uparrow\,\rangle_{j}$.

An eigenstate corresponds to a set of the spectral parameters,
$\{\lambda_{j}\}$, which is a solution of the eqs.~(\ref{eqn:BAE}).
The state vector can be expressed by using the operator of an
off-diagonal element of the monodromy matrix;
\begin{equation}
  \label{eigenvec}
  |\{\lambda_{j}\}\rangle
  = \prod_{j=1}^{M} [T(\lambda_{j})]_{1,2} |\,\emptyset\,\rangle.
\end{equation}

Now we calculate the eigenvalue of the $k$-th order
Hamiltonian~(\ref{H:Hk}) corresponding to the state (\ref{eigenvec}).
The eigenvalue of the transfer matrix is calculated as follows;
\begin{equation}
  \label{eqn:eigen:TM}
  \tau(\mu)|\{\lambda_{j}\}\rangle
  = \theta(\mu,\{\lambda_{j}\})|\{\lambda_{j}\}\rangle,
\end{equation}
where
\begin{equation}
  \label{eigen:TM}
  \theta(\mu,\{\lambda_{j}\})
  = (a(\mu))^N \prod_{j=1}^{M} f(\lambda_{j},\mu)
  + (d(\mu))^N \prod_{j=1}^{M} f(\mu,\lambda_{j}).
\end{equation}
Taking a logarithmic derivative of this equation, we obtain the
eigenvalue of the $k$-th order Hamiltonian $H^{(k)}$ for $k\ge1$;
\begin{equation}
  \label{eigen:Hk;0}
  E_{k}(\{\lambda_{j}\})
  =
  \left.
    \left( \frac{\sin\gamma}{\gamma}
           \frac{\partial}{\partial\mu} \right)^{k-1}
    E(\mu,\{\lambda_{j}\})
  \right|_{\mu=\im},
\end{equation}
with
\begin{equation}
  \label{def:E}
  E(\mu,\{\lambda_{j}\})
  =
  \frac{\im\sin\gamma}{\gamma}
  \frac{\partial}{\partial\mu}
  \ln\theta(\mu,\{\lambda_{j}\}).
\end{equation}
The second term on the {\rhs} of the eq.~(\ref{eigen:TM}) does not
contribute to the eigenvalue of the Hamiltonian if $k<N$, since the
function $(d(\mu))^N$ and its derivatives vanish at $\mu=\im$.  We
consider the thermodynamic limit, $N\to\infty$, and the term can be
omitted.  Then the function~(\ref{def:E}) reduces to 
\begin{equation}
  \label{func:E}
  E(\mu,\{\lambda_{j}\})
  = \sum_{j=1}^{M} e(\mu,\lambda_{j}),
\end{equation}
with
\begin{equation}
  \label{func:e}
  e(\mu,\lambda)
  = \frac{\sin^2\gamma}
         {\cos\gamma-\cosh\gamma(\lambda-\mu+\im)}.
\end{equation}
The eigenvalue of the $k$-th order Hamiltonian is given as the
derivative of the function~(\ref{func:E});
\begin{equation}
  \label{eigen:Hk;1}
  E_{k}(\{\lambda_{j}\})
  = \sum_{j=1}^{M} e_{k}(\lambda_{j}),
\end{equation}
with
\begin{equation}
  \label{energy:1p}
  e_{k}(\lambda)
  =
  \left(
    -\frac{\sin\gamma}{\gamma}
     \frac{\partial}{\partial\lambda}
  \right)^{k-1} e(\im,\lambda).
\end{equation}
Note that the following identity is used in the derivation of the
eq.~(\ref{energy:1p});
\begin{equation}
  \frac{\partial}{\partial\mu} e(\mu,\lambda_{j})
  =
  -\frac{\partial}{\partial\lambda_{j}} e(\mu,\lambda_{j}).
\end{equation}

The expression~(\ref{eigen:Hk;1}) means that the system is composed of
$M$ particles, each of which has rapidity $\lambda$, and the energy of
the system is given as the sum of one-particle
energies~(\ref{energy:1p}).

\section{Phase Diagram of the Single-Chain Model}

We determine the phase diagram of the system~(\ref{H:def}) here in the 
thermodynamic limit.  The energy for the eigenstates~(\ref{eigenvec})
is given from the results of the previous section;
\begin{equation}
  \label{energy}
  E = \sum_{j=1}^{M}
    \biggl[ 1+\alpha
      \left(
        \frac{\sinh\gamma}{\gamma}
        \frac{\partial}{\partial\lambda_{j}}
      \right)^{2}
    \biggr]
    \frac{\sin^{2}\gamma}{\cos\gamma-\cosh\gamma\lambda_{j}}.
\end{equation}
The parameter $\gamma$, which is related with the anisotropy as
$\Delta=\cos\gamma$, takes a real value when $|\Delta|\le1$ and a
purely imaginary value when $\Delta>1$.

\subsection{$|\Delta|<1$}

Here we consider the case that the absolute value of the anisotropy is
less than one.  The Bethe ansatz equations~(\ref{eqn:BAE}) have string
solutions~\cite{TakaSuzu,Taka97} in the thermodynamic limit
$N\to\infty$;
\begin{eqnarray}
  \lambda_{j}^k &=& \lambda_{j} + (n_{j}-2k+1)\im
  + \frac{1-v_{j}}{2\gamma}\pi\im,
  \nonumber\\
  && \mbox{}
  \;\;\;\;\;\;\;\;\;\;\;\;\;\;\;\;\;\;\;\;\;
  {\rm for}\;\; k=1,\dots,n_{j},
\end{eqnarray}
where the length, $n_{j}$, of a string is bounded by the number of the
particles, $\sum n_{j} = M$, and the parity, $v_{j}$, of a string takes 
value of $\pm1$.  The parities originate from the periodicity of the
eqs.~(\ref{eqn:BAE}) along the imaginary axis.  It is thought that the
string solutions are sufficient to form thermodynamics.

Following the usual procedure~\cite{TakaSuzu,Taka97}, we classify the
solutions of the Bethe ansatz equations by the lengths and the
parities of strings.  The density of rapidity and that of holes,
$\rho_{j}(\lambda)$ and $\rho_{j}^{\rm h}(\lambda)$, where the
subscript denotes a class, is defined.  In the thermodynamic Bethe
ansatz method, the dressed energy, $\varepsilon_{j}(\lambda)$, is
introduced as
\begin{equation}
  \label{def:dress}
  \varepsilon_{j}(\lambda) = T\ln \eta_{j}(\lambda),
  \;\;\;
  \eta_{j}(\lambda)
  = \frac{\rho_{j}^{\rm h}(\lambda)}{\rho_{j}(\lambda)}.
\end{equation}
We can deduce a set of functional equations about $\rho_{j}(\lambda)$,
$\rho_{j}^{\rm h}(\lambda)$, and $\eta_{j}(\lambda)$ under the
thermodynamic equilibrium condition, but it is lengthy and cumbersome
to write down them and they are omitted here. (See
refs.~\citen{TakaSuzu} and \citen{Taka97} if necessary.)

Though the number of the classes is infinite, in the low temperature
limit we only have to consider two of them, both of which belong to
the length one, one belongs to the positive parity and the other to
the negative, since it is shown by Takahashi and
Suzuki~\cite{TakaSuzu} that the energies of the classes belonging to
the lengths longer than one are large and that they don't contribute
to the ground state.  Then we obtain the following functional
equations about the densities of rapidity, $\rho_{\pm}(\lambda)$ and
$\rho_{\pm}^{\rm h}(\lambda)$, and the dressed energies,
$\varepsilon_{\pm}(\lambda)$, of the two classes;
\begin{eqnarray}
  \label{eqn:rho0}
  g_{\pm}(\lambda)
  &=&
  \pm\left(
        \rho_{\pm}(\lambda) + \rho_{\pm}^{\rm h}(\lambda)
     \right)
  \nonumber\\
  && \mbox{}
  + K_{\pm}\!*\rho_{+}(\lambda)
  + K_{\mp}\!*\rho_{-}(\lambda),
  \bigskip
  \\
  \label{eqn:dress0}
  \varepsilon_{\pm}(\lambda)
  &=&
  \varepsilon^0_{\pm}(\lambda)
  - K_{\pm}\!*\varepsilon_{+}^{(-)}(\lambda)
  + K_{\mp}\!*\varepsilon_{-}^{(-)}(\lambda),
\end{eqnarray}
with
\begin{eqnarray*}
  g_{\pm}(\lambda)
  &=& \pm\frac{1}{2\pi}
      \frac{\gamma\sin\gamma}{\cosh\gamma\lambda\mp\cos\gamma},
  \\
  K_{\pm}(\lambda)
  &=& \pm\frac{1}{2\pi}
  \frac{\gamma\sin 2\gamma}{\cosh\gamma\lambda\mp\cos 2\gamma},
  \\
  \varepsilon^0_{\pm}(\lambda)
  &=& -\frac{2\pi\sin\gamma}{\gamma}
  \biggl[ 1+\alpha
    \left( \frac{\sin\gamma}{\gamma}
           \frac{\partial}{\partial\lambda} \right)^{2}
  \biggr] g_{\pm}(\lambda),
  \\
  \varepsilon_{\pm}^{(-)}(\lambda)
  &=&
  \left\{
    \begin{array}{lc}
      \varepsilon_{\pm}(\lambda) & \varepsilon_{\pm}(\lambda)\le0,
      \medskip\\
      0 & {\rm otherwise},
    \end{array}
  \right.
\end{eqnarray*}
and the asterisk product $a*b(\lambda)$ denotes a convolution defined
as
\begin{equation}
  a*b(\lambda)
  =
  \int_{-\infty}^{\infty} a(\lambda-\mu)\, b(\mu) \D\mu.
\end{equation}
Note that  in the low temperature limit, $T\to0$, the following
relation holds from the definition~(\ref{def:dress}): the rapidity
density $\rho(\lambda)$ (the hole density $\rho^{\rm h}(\lambda)$) is
zero when the dressed energy takes a positive (negative) value.  The
eqs.~(\ref{eqn:rho0}) together with this relation become complete to
determine the densities of rapidity.

The analytical solutions of the dressed energies are easily obtained
if the coupling constant $|\alpha|$ is sufficiently small.  By
applying the Fourier transformation to the eqs.~(\ref{eqn:dress0}) we
obtain
\begin{equation}
  \label{func:dress0}
  \varepsilon_{\pm}(\lambda)
  =
  \mp\frac{\pi \sin\gamma}{2\gamma}
  \biggl[ 1+\alpha
    \left(
      \frac{\sin\gamma}{\gamma}
      \frac{\partial}{\partial\lambda}
    \right)^{2}
  \biggr]
  f_{\pm}(\lambda),
\end{equation}
with
\begin{eqnarray*}
  f_{+}(\lambda)
  &=&
  \sech \frac{\pi\lambda}{2},
  \\
  f_{-}(\lambda)
  &=&
  \left\{
  \begin{array}{lc}
    0 & \displaystyle{0<\gamma\le\frac{\pi}{2}},
    \smallskip\\
    \displaystyle{\frac{-\cos(\pi^{2}/2\gamma)\cosh(\pi\lambda/2)}
                       {\cosh\pi\lambda+\cos(\pi^{2}/\gamma)}} &
    \displaystyle{\frac{\pi}{2}<\gamma<\pi}.
  \end{array}
  \right.
\end{eqnarray*}
The solutions~(\ref{func:dress0}) are valid as long as
$\varepsilon_{+}(\lambda) \le 0$ and $\varepsilon_{-}(\lambda) \ge 0$
for arbitrary value of the rapidity $\lambda$.  (Note that
$-\cos(\pi^{2}/2\gamma)$ is positive when $\pi/2 < \gamma < \pi$.)  It
is easy to see from the solutions that this condition is satisfied
when the coupling constant lies in the finite range, $\alpha_{\rm
  crit}^{-} < \alpha < \alpha_{\rm crit}^{+}$, where
\begin{equation}
  \label{crit0}
  \alpha_{\rm crit}^{\pm}
  =
  \pm\left( \frac{2\gamma}{\pi\sin\gamma} \right)^{2}
  h_\gamma^\pm(\pi^{2}/\gamma),
\end{equation}
with
\begin{eqnarray*}
  h_\gamma^{+}(x)
  &=&
  \left\{
  \begin{array}{lc}
    1 & \displaystyle{0<\gamma\le\frac{\pi}{2}},
    \smallskip\\
    \displaystyle{\frac{1-\sin^{2}(x/2)}{1+\sin^{2}(x/2)}}
    & \displaystyle{\frac{\pi}{2}<\gamma<\pi},
  \end{array}
  \right.
  \\
  h_\gamma^{-}(x)
  &=&
  \left\{
  \begin{array}{lc}
    1 & \displaystyle{0<\gamma\le\frac{3\pi}{5}},
    \smallskip\\
    \displaystyle{\frac{2\sin^{2} x}{2\cos^{2}x-4\cos x+3}}
    & \displaystyle{\frac{3\pi}{5}<\gamma<\pi}.
  \end{array}
  \right.
\end{eqnarray*}

The ground state is the same as that of the {\it XXZ} model when the
coupling constant is in this range, since the Fermi points of the
dressed energies are the same.  When the coupling constant goes beyond 
the range, there add new four Fermi points.

The appearance of the new Fermi points is easily observed at the `free 
fermion point', $\Delta=0$, where the {\it XXZ} model reduces to a
free fermion system.  The one-particle dispersion of our model is
easily found to be
\begin{equation}
  \label{func:disp0}
  \varepsilon(p)
  =
  (1-\frac{\alpha}{2}) \cos{p}
  - \frac{\alpha}{2} \cos{3p}.
\end{equation}
There appear three negative-energy regions with six Fermi points when
the coupling constant is beyond the critical value, $|\alpha| >
\alpha_{\rm crit}^{\pm} = 1$.

Similar situation can be observed on the two dressed energies at
another value of the anisotropy parameter.  Then we conclude that
there occurs a phase transition if the coupling constant goes across
the critical value~(\ref{crit0}).  The new phase is characterized by
added new branches of excitations, which are also gapless.

The ground state in the new phase has no magnetic moment.  We can see
this fact by taking a Fourier component of one of the
eqs.~(\ref{eqn:rho0}) at zero frequency.  It leads to
\begin{eqnarray}
  \label{mag0}
  \!\!\!\!\!\!\!\!\!\!
  M/N &=& S^{z}_{\rm total}/N
  \nonumber\\
  &=& \frac{1}{2}
  - \int \rho_{+}(\lambda) \D\lambda
  - \int \rho_{-}(\lambda) \D\lambda
  \nonumber\\
  &=&
  \left\{
  \begin{array}{lc}
    \displaystyle{\frac{\pi}{2\gamma}
      \int \rho_{-}^{\rm h}(\lambda) \D\lambda}
    & \displaystyle{0<\gamma\le\frac{\pi}{2}},
    \medskip\\
    \displaystyle{\frac{1}{2(1\!-\!\gamma/\pi)}
        \int \rho_{+}^{\rm h}(\lambda) \D\lambda}
    & \displaystyle{\frac{\pi}{2}<\gamma<\pi}.
  \end{array}
  \right.
\end{eqnarray}
Since it can be proved that $\varepsilon_{-(+)}(\lambda) \le 0$
everywhere for $0 < \gamma \le \pi/2$ ($\pi/2 < \gamma < \pi$), there
is no hole of rapidity belonging to $-$($+$)-parity.  The ground state 
is, therefore, not magnetized at any strength of the
interaction~(\ref{H:int}).  The new phase is similar to the
corresponding one of the Tsvelik-Frahm model~\cite{Tsvelik,Frahm} in
this point.

\subsection{$\Delta\ge1$}

Here the parameter $\gamma$ is purely imaginary and we put $\gamma =
\im\theta$, $\,\theta\ge0$.  Since in this case there is no
periodicity along the imaginary axis in the Bethe ansatz
equations~(\ref{eqn:BAE}), only string solutions with positive parity,
which are centered on the real axis, are allowed.  On the other hand,
the equations become periodic along the real axis.  We restrict,
therefore, the rapidities in the following range;
\begin{equation}
  \label{def:Lambda}
  \Lambda = \{\lambda \,| -\frac{\pi}{\theta} \le \lambda \le
  \frac{\pi}{\theta}\}.
\end{equation}
(When the parameter $\gamma$ is zero, an adequate limit, $\theta\to0$,
should be taken on the equations below in this section and the 
range of the rapidity should be expanded to infinity; $\Lambda =
\{\lambda\,|-\infty<\lambda<\infty\}$.)

We only have to consider string solutions belonging to the length one 
in the same way as the previous section.  The density of rapidity,
$\rho(\lambda)$, and the dressed energy, $\varepsilon(\lambda)$, are
determined by the following eqs.~\cite{Taka,Gaudin,Taka97};
\begin{eqnarray}
  \label{eqn:rho1}
  g(\lambda)
  &=& \rho(\lambda) + \rho^{\rm h}(\lambda) + K\!*\rho(\lambda),
  \\
  \label{eqn:dress1}
  \varepsilon(\lambda)
  &=& \varepsilon^{0}(\lambda) - K\!*\varepsilon^{(-)}(\lambda),
\end{eqnarray}
with
\begin{eqnarray*}
  g(\lambda)
  &=&
  \frac{1}{2\pi}
  \frac{\theta \sinh\theta}{\cosh\theta - \cos\theta\lambda},
  \\
  K(\lambda)
  &=&
  \frac{1}{2\pi}
  \frac{\theta \sinh 2\theta}{\cosh 2\theta - \cos\theta\lambda},
  \\
  \varepsilon^{0}(\lambda)
  &=&
  -\frac{2\pi \sinh\theta}{\theta}
  \biggl[ 1+\alpha
    \left( \frac{\sinh\theta}{\theta}
           \frac{\partial}{\partial\lambda}
    \right)^{2}
  \biggr] g(\lambda),
  \\
  \varepsilon^{(-)}(\lambda)
  &=&
  \left\{
    \begin{array}{lc}
      \varepsilon(\lambda) & \varepsilon(\lambda)\le0,
      \medskip\\
      0 & {\rm otherwise},
    \end{array}
  \right.
\end{eqnarray*}
and the asterisk product $a*b(\lambda)$ denotes a convolution over the
range~(\ref{def:Lambda}) here;
\begin{equation}
  a*b(\lambda)
  =
  \int_{\Lambda} a(\lambda-\mu)\,b(\mu) \D\mu.
\end{equation}

The eq.~(\ref{eqn:dress1}) has an analytical solution for sufficiently
small $|\alpha|$.  The dressed energy is obtained as
\begin{eqnarray}
  \label{func:dress1}
  \varepsilon(\lambda)
  &=&
  -\biggl[ 1+\alpha
     \left(
       \frac{\sinh\theta}{\theta}
       \frac{\partial}{\partial\lambda}
     \right)^{2}
   \biggr]
  \sum_{n=-\infty}^{+\infty}
  \frac{1}{2} \sech |n|\theta
  \nonumber\\
  &=&
  -\frac{K(k)\sinh\theta}{\pi}
  \biggl[ 1+\alpha
    \left(
      \frac{\sinh\theta}{\theta}
      \frac{\partial}{\partial\lambda}
    \right)^{2}
  \biggr]
  \nonumber\\
  &&
  \times \dn(K'(k)\lambda;k),
\end{eqnarray}
where the parameters $K(k)$ and $K'(k)$ are the complete elliptic
integrals of the first kind with modulus $k$ and $k'= \sqrt{1-k^{2}}$,
respectively, and the function $\dn(x;k)$ is one of Jacobi's elliptic
function with modulus $k$.  The modulus is related with the anisotropy
as
\begin{equation}
  \label{modulus}
  \Delta = \cosh\theta,\;\;\;
  \theta = \pi\frac{K'(k)}{K(k)}.
\end{equation}

The solution~(\ref{func:dress1}) is valid as long as
$\varepsilon(\lambda) \le 0$ everywhere.  This condition is identical
with the following one;
\begin{equation}
  \label{crit1}
  |\alpha| < \alpha_{\rm crit}
  = \left( \frac{\theta}{k K'(k) \sinh\theta} \right)^{2}.
\end{equation}
The critical value is continuous at $\gamma=0$ with that of the
previous section taking value of $\pm(2/\pi)^2$.

The ground state coincides with that of the {\it XXZ} model when the
condition~(\ref{crit1}) is satisfied.  It is not magnetized.  If the
anisotropy is in the `Ising-like region', $\Delta>1$, the ground
state also shows the N\'eel order and has excitations with gap.

When the coupling constant is beyond the critical value, the dressed
energy gains a positive region where $\varepsilon(\lambda) > 0$.  We
name the region $D^{+}$, and define its complimentary set $D^{-}$;
\begin{equation}
  \label{def:D}
  D^{+} = \{ \lambda\in\Lambda\,|\: \varepsilon(\lambda)>0 \},
  \;\;\;
  D^{-} = \Lambda - D^{+}.
\end{equation}
The existence of not-empty region $D^{+}$ means that there appear new
Fermi points and the system has gapless excitations, since the dressed
energy function, $\varepsilon(\lambda)$, is continuous and takes a
definite value at the end points of domain $\Lambda$.  Thus the system
shows a phase transition at the critical value of the coupling
constant.  When the anisotropy parameter lies in the `Ising-like
region', the system undergoes a transition from the gapful phase to
the gapless one when the coupling constant exceeds the critical value.

Another typical character of the new phase is that the ground state is
magnetized.  We can calculate the magnetic moment as
\begin{eqnarray}
  \label{mag1}
  M/N &=& S_{\rm total}^{z}/N
  \nonumber\\
  &=& \frac{1}{2} - \int_{\Lambda} \rho(\lambda) \D\lambda
  \nonumber\\
  &=& \frac{1}{2} \int_{\Lambda} \rho^{\rm h}(\lambda) \D\lambda,
\end{eqnarray}
by taking the zero frequency component of the Fourier expansion of the
eq.~(\ref{eqn:rho1}).  As we noted in the previous section, the hole
density, $\rho^{\rm h}(\lambda)$, can have non zero value only in the
region $D^{+}$.  When the coupling constant is large enough for the
region $D^{+}$ not to be empty, it is easy to make sure from the
eq.~(\ref{eqn:rho1}) that there exist holes.  Thus the last term of
the eq.~(\ref{mag1}) gives a positive value and the system obtains a
magnetic moment.

Note that the system is not completely but partly magnetized even
for quite large coupling constant, since the region $D^{-}$ never
becomes empty.  In fact the magnetic moment gets saturated at
$S^{z}_{\rm total}/N=0.3217$ ($0.1479$) when $\alpha\to+\infty$
($-\infty$) and $\Delta=1$.

\begin{figure}
  \epsfxsize=3.3in
  \leavevmode
  \epsfbox{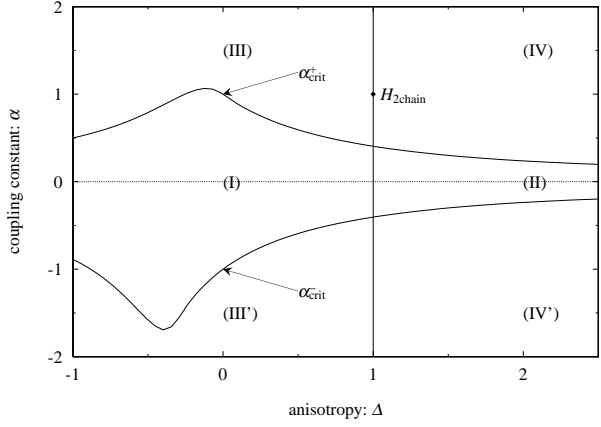}
 \caption{Phase diagram of the single-chain model~(\ref{H:def}) is
    given.  Its ground state has the same property as that of the {\it
      XXZ} model for sufficiently small $|\alpha|$ between
    $\alpha_{\rm crit}^{+}$ and $\alpha_{\rm crit}^{-}$; the phase
    (I), $\Delta\le1$, has gapless excitations, the phase (II),
    $\Delta>1$, has excitations with gap and the ground states in both
    phases are not magnetized.  The systems in the region (III) and (IV)
    obtain additional branches of excitation to those in phase (I) and
    (II), respectively.  The extra excitations in both phases are
    gapless ones.  The ground state in the phase (III) is also not
    magnetized, though in phase (IV), $\Delta\ge1$, it obtains a
    finite, but not-full, magnetization.  When the parameters are
    chosen as $\Delta=\alpha=1$, the model reduces to the two-chain
    model, which lies in the partially magnetized phase (IV).}
  \label{fig:phase}
\end{figure}

\section{Two-Chain Model}

Now we can get information of the two-chain model.  We only have to
watch the single-chain model~(\ref{H:def}) at a special value of
the parameter, $\alpha=1$ and $\Delta=1$.  From the view point of the
single-chain model, the coupling constant is beyond the critical value,
$\alpha_{\rm crit} = (2/\pi)^{2} < 1$.  Then the system of the
two-chain model stays in the partly magnetized phase.

The magnetic moment is calculated numerically;
\begin{equation}
  \label{mag2ch}
  S^{z}_{\rm total}/N = 0.1049,
\end{equation}
which shows that the system is about 20\% magnetized in the ground
state.  As noted in the previous section, the system has gapless
excitations with extra branches.

\section{Twisted Four-Body Correlation Function}

The expectation value of the $k$-th order Hamiltonian operator,
$H^{(k)}$, for the ground state of the isotropic and
anti-ferromagnetic Heisenberg model ({\it XXX} model) is presented in
this section.  We evaluate it by using the eq.~(\ref{eigen:Hk;1}) and
the rapidity distribution function, $\rho_{0}(\lambda)$, of the ground
state of the {\it XXX} model;
\begin{eqnarray}
  \label{func:rho0}
  \rho_{0}(\lambda) &=& \frac{1}{4} \sech\frac{\pi\lambda}{2}
  \nonumber\\
  &=& \frac{1}{4\pi} \int_{-\infty}^{\infty}
  \sech\,\omega \exp(-\im\lambda\omega) \D\omega.
\end{eqnarray}
Note that in this section we analyze only the isotropic case,
$\Delta=1$.

The expectation value, $\langle H^{(k)}\rangle_{0}$, is calculated from
the eq.~(\ref{eigen:Hk;1}) in an analytical form;
\begin{equation}
  \label{expect:Hk}
  E_{k} = \frac{\langle H^{(k)} \rangle_{0}}{N}.
\end{equation}
This vanishes if the integer $k$ is even, and it gives the following
value for odd $k$;
\begin{eqnarray}
  E_{k}
  &=&
  \int_{-\infty}^{\infty}
  e_{k}(\lambda)\, \rho_{0}(\lambda) \D\lambda
  \nonumber\\
  &=&
  \left\{
    \begin{array}{l}
      (-1)^{(k+1)/2} 2^{-k+1}(1-2^{-k+1})\\
      \begin{array}{lc}
        \times (k-1)!\, \zeta(k) & {\rm for}\; k\ge3,
        \medskip\\
        \!\!\!-\ln2 & {\rm for}\; k=1,
      \end{array}
    \end{array}
  \right.
\end{eqnarray}
where the function $\zeta(s)$ is Riemann's zeta, $\zeta(s) =
\sum_{n=0}^{\infty}n^{-s}$.

The expectation value of the first order Hamiltonian leads the nearest 
neighbor correlation function of the {\it XXX} model;
\begin{eqnarray}
  \label{expect:HXXZ}
  \langle \mib{S}_{j}\cdot\mib{S}_{j+1} \rangle_{0}
  &=&
  \frac{\langle H^{(0)} \rangle_{0}}{N} + \frac{1}{4}
  \nonumber\\
  &=&
  \frac{1}{4} - \ln2.
\end{eqnarray}

\begin{figure}
  \epsfxsize=3.3in
  \leavevmode
  \epsfbox{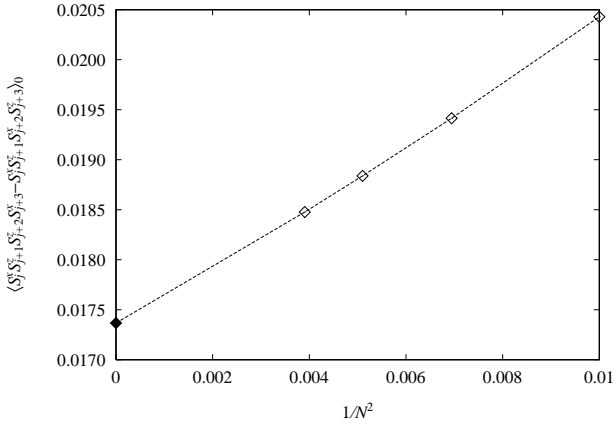}
  \caption{The twisted four-body correlation functions of the
   anti-ferromagnetic Heisenberg chain with $N$ sites, $N=10,12,14,$
    and $16$, are plotted as a function of $1/N^2$.  In the
    thermodynamic limit, $N\to\infty$, these values approach the
    theoretical value (\ref{corr:4body}).}
  \label{fig:4corr}
\end{figure}

Similarly the expectation value of the two-chain model
Hamiltonian~(\ref{H:11}) gives the following relation;
\begin{eqnarray}
  \label{eqn:corr}
  &&
  \frac{1}{2}
  \langle \mib{S}_{j}\cdot\mib{S}_{j+2} \rangle_{0}
  + 2\langle
       (\mib{S}_{j  }\!\times\!\mib{S}_{j+1})\cdot
       (\mib{S}_{j+2}\!\times\!\mib{S}_{j+3})
     \rangle_{0}
  \nonumber\\
  && = \frac{1}{8} - \ln2 + \frac{3}{8}\zeta(3).
\end{eqnarray}
The correlation function between the second-neighbor sites is
obtained by one of the authors~\cite{Taka77};
\begin{equation}
  \label{expect:NNN}
  \langle \mib{S}_{j} \cdot \mib{S}_{j+2} \rangle_{0}
  = \frac{1}{4} - 4\ln2 + \frac{9}{4} \zeta(3).
\end{equation}
Then the twisted four-body correlation function is calculated as
\begin{eqnarray}
  \label{expect:4body}
  \langle
    (\mib{S}_{j  }\!\times\!\mib{S}_{j+1})\cdot
    (\mib{S}_{j+2}\!\times\!\mib{S}_{j+3})
  \rangle_{0}
  &=& \frac{1}{2}\ln2 - \frac{3}{8}\zeta(3)
  \nonumber\\
  &=& -0.104198,
\end{eqnarray}
or expressed by its component as
\begin{eqnarray}
  \label{corr:4body}
  &&
  \langle
    S^{\alpha}_{j}S^{\beta}_{j+1}S^{\beta }_{j+2}S^{\alpha}_{j+3}
  - S^{\alpha}_{j}S^{\beta}_{j+1}S^{\alpha}_{j+2}S^{\beta }_{j+3}
  \rangle_{0}
  \nonumber\\
  && =
  \left(
    \frac{1}{16}\zeta(3) - \frac{1}{12}\ln2
  \right)
  (1-\delta_{\alpha\beta})
  \nonumber\\
  && = 0.0173663\,(1-\delta_{\alpha\beta}),
  \;\;\;
  \alpha,\beta = x,y,z.
\end{eqnarray}
To check the validity of this result, we calculate the four-body
correlation function for finite systems using the exact
diagonalization method and plot them.  Figure.~\ref{fig:4corr}
supports that the eq.~(\ref{corr:4body}) is correct.

\section{Summary}

The {\it XXZ} model with competing four-body interaction has been
studied.  When the four-body interaction is weaker than the critical
value, $\alpha_{\rm crit}$, the ground state coincides with that of
the {\it XXZ} model and the spectrum of low-lying excitation is also
the same.  If it gets stronger than $\alpha_{\rm crit}$, the new
phases arise.  The critical value of the coupling constant has been
obtained analytically.  The new phases (III) and (IV), see
Fig.~\ref{fig:phase}, have gapless excitations.  The phase (III),
$|\Delta|<1$, the ground state is not magnetized and the phase (IV),
$\Delta\ge1$, it is magnetized in part.

The two-chain model~(\ref{H:2chain}) coupled by four-body interactions
can be deduced from the above single-chain model.  Since the former
model corresponds to the latter in the phase (IV), the ground state of
the two-chain model shows the partly magnetized property and has 
excitations without gap.

From the expectation value of the four-body interaction operator, the
twisted four-body correlation of the anti-ferromagnetic Heisenberg
model has also been obtained analytically.

\section*{Acknowledgment}

We would like to thank T.~Kawarabayashi, K.~Itoh, and M.~Nakamura for
helpful comments and discussions.  To check our result we have used
the computer program TITPACK Ver. 2 developed by Professor
H.~Nishimori, to whom we are indebted.

\end{document}